\documentstyle[12pt]{article}
\newcommand{\newsection}{    
\setcounter{equation}{0}
\section}
\def\appendix#1{
\addtocounter{section}{1}
\setcounter{equation}{0}
\renewcommand{\thesection}{\Alph{section}}
\section*{Appendix \thesection\protect\indent #1}
\addcontentsline{toc}{section}{Appendix \thesection\ \ \ #1}
}
\newcommand{\rf}[1]{(\ref{#1})}

\def\be{\begin{equation}}
\def\ee{\end{equation}}
\newcommand{\beq}{\begin{equation}}
\newcommand{\eeq}{\end{equation}}
\newcommand{\bea}{\begin{eqnarray}}
\newcommand{\eea}{\end{eqnarray}}

\renewcommand{\a}{\alpha}

\newcommand{\om}{\omega}

\newcommand{\oh}{\frac{1}{2}}

\newcommand{\tr}{{\,\rm Tr}\;}

\newcommand{\cim}{\oint_{C}\frac{d\omega}{2\pi i}}

\newcommand{\ci}{\oint_{C}\frac{d\omega}{2\pi i}\;\frac{
V'(\omega)}{p-\omega}}

\newcommand{\LA}{\left\langle}
\newcommand{\RA}{\right\rangle}

\def\e{{\,\rm e}\,}

\begin{document}
\topmargin 0pt
\oddsidemargin 5mm
\headheight 0pt
\headsep 0pt
\topskip 9mm

\hfill    NBI-HE-93-30

\hfill May 1993

\addtolength{\baselineskip}{0.20\baselineskip}

\begin{center}

\vspace{24pt}
{\large \bf
{}From 1-matrix model to Kontsevich model}

\vspace{24pt}

{\sl J. Ambj\o rn and C.F.\ Kristjansen}

\vspace{6pt}

 The Niels Bohr Institute\\
Blegdamsvej 17, DK-2100 Copenhagen \O , Denmark\\

\end{center}
\vspace{18pt}

\begin{center}
{\bf Abstract}
\end{center}

Loop equations of matrix models express the invariance of the models
under field redefinitions. We use loop equations to prove that it is
possible to
define continuum times for the generic
hermitian \mbox{1-matrix} model such that
all correlation functions in the double scaling limit
agree with the corresponding correlation functions of the Kontsevich
model expressed in terms of kdV times. In addition the double scaling
limit of the partition function of the hermitian matrix model agree with
the $\tau$-function of the kdV hierarchy corresponding to the
Kontsevich model ({\it and not the square of the $\tau$-function})
except for some complications at genus zero.

\newpage

\newsection{Introduction}

Since the discovery of the Kontsevich model as a matrix model
realization of a $\tau$-function of the kdV hierarchy~\cite{Kon91}
there has
been attempts to relate the partition function of this model to the
continuum partition function of various 1-matrix models. In
reference~\cite{KMMM} a limiting procedure which allows one, on path
integral level, to pass from the partition function of the reduced
hermitian matrix model to the {\it square} of the partition function of the
Kontsevich model was presented. However this procedure relied very
heavily on the assumption of a {\it symmetrical} potential for the 1-matrix
model. In reference~\cite{hermitian} a different limiting procedure
provided a way of passing from the partition function of the generic
hermitian 1-matrix model to the {\it non squared} partition
function of the Kontsevich model. However the latter limiting procedure
was rather unconventional, involving an analytic continuation of the
size of the matrix of the original 1-matrix model from $N$ to $-\xi
N$.

Here we will show using the conventional double scaling limit
how one can define continuum times for the generic
hermitian matrix model so that
its continuum
partition function apart from some complications at genus zero
turn directly into the {\it non
squared} partition function of the Kontsevich
model. Our proof is based on the loop equations of the two models and is
facilitated by the use of the so called moment description of the
1-matrix model introduced in reference~\cite{hermitian}. (See
also~\cite{ACM} and~\cite{complex} for earlier versions.)

The content of
section~\ref{1mm} is a short review of the moment description
away from and in the double scaling limit. In section~\ref{Kontm} we have
collected some useful formula for the Kontsevich model.
Section~\ref{tau} contains a motivation for our choice of continuum
time variables, a derivation of an appropriate version of the loop
equations for the hermitian matrix model as well as for the Kontsevich
model and finally the proof of our statement. In section~\ref{discussion}
we discuss how our procedure can be applied to the complex matrix model
and the symmetrical hermitian one. In particular we show that there is
no contradiction between the results of this paper and those of
references~\cite{KMMM}, \cite{MMMM} and~\cite{FKN}.

\newsection{The hermitian 1-matrix model \label{1mm} }
\subsection{Discrete description}
The hermitian 1-matrix model is defined by the partition function
\beq
Z[\{g_j\},N]=e^{N^2 F}=\int_{N\times N}
d\phi\exp(-N\tr V( \phi)) \label{partition}
\eeq
where the integration is over hermitian $N\times N$ matrices and
\beq
V(\phi)=\sum_{j=1}^{\infty}\frac{g_j}{j}\,\:\phi^j
\label{potential}
\eeq
If $g_{2k+1}=0$ for all values of $k$ we will denote the model as
symmetrical. Otherwise we will denote it as generic.
The observables of the model are the s-loop correlators
\beq
W(p_1,\ldots,p_s)=N^{s-2} \LA \tr\frac{1}{p_1-\phi} \ldots
\tr\frac{1}{p_s-\phi} \RA_{conn.}
\label{Wconn}
\eeq
where the expectation value is defined in the conventional way and
where $conn$ refers to the connected part. The multi-loop correlators can
be found from the free energy
by application of the loop insertion operator,
$\frac{d}{dV(p)}$:
\beq
W(p_1,\ldots, p_s)=\frac{d}{dV(p_s)}\;\frac{d}{dV(p_{s-1})}\ldots
\frac{d}{dV(p_1)}F
\label{loopaverage}
\eeq
where
\beq
\frac{d}{dV(p)}\equiv
-\sum_{j=1}^{\infty}\frac{j}{p^{j+1}}\frac{d}{dg_{j}}\;
\eeq
With the normalization chosen above, the genus expansion of the
correlators reads
\beq
W(p_1,\ldots,p_s)=\sum_{g=0}^{\infty}\frac{1}{N^{2g}}\;
W_{g}(p_1,\ldots,p_s) \hspace{1.0cm} (s\geq 1)\,. \label{Wgen}
\eeq
Similarly we have
\beq
F=\sum_{g=0}^{\infty}\frac{1}{N^{2g}}F_g
\label{genexp}
\eeq
The genus $g$ contribution to the 1-loop correlator can be found by
solving by iteration the loop equations of the model.
These equations express the invariance of~\rf{partition} under field
redefinitions of the form
\beq
\phi \rightarrow \phi+\epsilon\sum_{n\geq0}\,\frac{\phi^n}{p^{n+1}}
=\phi+\epsilon\frac{1}{p-\phi}
\eeq
Under a transformation of this type the measure changes as
\beq
d\phi\ \rightarrow d\phi
\left(1+\epsilon\left[\tr\left(\frac{1}{p-\phi}\right)\right]^2\right)
\eeq
and we get to first order in $\epsilon$
\beq
\int d\phi\left\{\left(\tr\left(\frac{1}{p-\phi}\right)\right)^2
-N\tr\left(V'(\phi)\,\,\frac{1}{p-\phi}\right)\right\}
e^{-N\tr V(\phi)} =0
\eeq
or introducing~\rf{Wconn}
\beq
\ci W(\om)= \left(W(p)\right)^2+\frac{1}{N^2} W(p,p)
\label{loopeqn}
\eeq
where $V(\om)=\sum_j g_j \om^j /j$ and $C$ is a curve
which encloses all possible eigenvalues of $\phi$ i.\e.\ all
singularities of $W(\om)$.
Here we will assume that the density of eigenvalues of
$\phi$ has support only on the interval $[y,x]$ and is normalized to
1. Then the singularities of
$W(p)$ consist of only one square root branch cut on the
real axis, $[y,x]$ and $W(p)$ behaves as $1/p$ as $p \rightarrow
\infty$~\cite{Brezin}.
With this assumption the genus-0 contribution to the one loop
correlator can be written as
\bea
W_{0}(p)&=&\oh \ci \left\{\frac{(p-x)(p-y)}{(\om-x)(\om-y)}\right\}^{1/2}
\label{W0}
\eea
where $x$ and $y$ are determined by the following boundary equations
\bea
B_1(x,y) &=&\cim\; \frac{V'(\om)}{\sqrt{(\om-x)(\om-y)}}=0\,,
\label{B1}
\\
B_2(x,y)&=&\cim\; \frac{\om V'(\om)}{\sqrt{(\om-x)(\om-y)}}=2\,.
 \label{B2}
\eea
{}From
the 1-loop correlator all multi-loop correlators
can be found by application of the loop insertion operator and the free
energy by application of the inverted loop insertion operator.
The higher genera contributions are most easily expressed by introducing
in stead of the couplings, $\{g_i\}$, some moments,
$\{M_i,J_i\}$~\cite{hermitian}. One of the advantages of this change of
variables is that the genus $g$ contribution to the $s$-loop correlator,
$W_g(p_1,\ldots,p_s)$, depends only on a finite number of moments namely
at most $2\times(3g-2+s)$ for the generic hermitian matrix model.
As opposed to this
$W_g(p_1,\ldots,p_s)$ depends on the entire set of coupling constants,
$\{g_i\}$. Furthermore the description in terms of moments
allows one to access very easily the continuum limit. For the hermitian
matrix model the moments are defined by
\bea
M_{k}(x,y,\{g_i\})&=&\cim\, \frac{V'(\om)}{(\om-x)^{k+1/2}\,(\om-y)^{1/2}}
\hspace{1.7cm} k\geq -1\,,
\label{moment1}\\
J_{k}(x,y,\{g_i\})&=&\cim\, \frac{V'(\om)}{(\om-x)^{1/2}\,(\om-y)^{k+1/2}}
   \hspace{1.7cm} k\geq -1 \,.\label{moment2}
\eea
Normally we would think of $x$ and $y$ as being fixed by the boundary
conditions~\rf{B1} and~\rf{B2} for given values of the coupling
constants. However it will prove convenient for the following to
consider $M_k$ and $J_k$ as functions of $x$ and $y$ as well as of the
coupling constants.
Furthermore we shall make use of the following rewriting of $W_0(p)$
\beq
W_0(p)=\frac{1}{2}V'(p)-
\frac{1}{4}(p-x)^{1/2}(p-y)^{1/2}
\sum_{q=1}^{\infty}\left\{(p-x)^{q-1}M_q+(p-y)^{q-1}J_q\right\}
\label{W0exp}
\eeq
It is derived by deforming the contour of integration in~\rf{W0} into
two, one which encloses the point $\om=p$ and one which encircles
infinity. To find the contribution from the latter we write
$(p-\om)^{-1}$ as
\beq
\frac{1}{p-\om}=\frac{1}{2}\left\{\frac{1}{(p-x)-(\om-x)}\right\}+
\frac{1}{2}\left\{\frac{1}{(p-y)-(\om-y)}\right\}
\eeq
and expand i powers of $\left(\frac{p-x}{\om-x}\right)$ and
$\left(\frac{p-y}{\om-y}\right)$ respectively. The expansion procedure
is justified by the fact that the contour of integration encircles
infinity which allows us to assume that $\om >>p$. We note that the two
bracketed terms in~\rf{W0exp} are actually identical. This is an example
of another appealing feature of the moment description. All expressions
are invariant under the interchanges $x\leftrightarrow y$,
$J\leftrightarrow M$ which of course just reflects the fact that a
priori there is nothing which allows us to distinguish between the two
endpoints of the cut. (In the continuum limit, which we are going to
consider, this will not be true.) Furthermore it is worth noting that the
expression~\rf{W0exp} allows us to determine very easily the inverse
transformations $g_i=g_i(x,y,\{M_i\},\{J_j\})$ since we know that
$W_0(p)$ only contains negative powers of $p$.
In reference~\cite{hermitian} explicit expressions
for $W_g(p)$ and $F_g$ in terms of
the moments were presented for the lowest genera. Furthermore their
general form were conjectured and proven by induction. Here we will only
be interested in terms which are relevant for the double scaling limit.

\subsection{The continuum limit}

Certain points in the (infinite dimensional)
coupling constant space are of particular
interest, namely the Kazakov multi-critical points~\cite{kaz}.
 The $m$'th multi-critical
points are in the case of a generic hermitian matrix model characterized
by the fact that the eigenvalue distribution which under normal
circumstances vanishes  as a square root at both endpoints of its
support acquires $(m-1)$ additional zeros at one end point, say $x$. The
condition for being at a $m$'th multi-critical point is in this case
\beq
M_1=M_2=\ldots= M_{m-1}=0,\hspace{0.4cm}M_m \neq 0,
\hspace{0.4cm}J_1\neq 0
\label{mcrit}
\eeq
as is seen from~\rf{W0exp}.
In other words the $m$'th multi-critical points constitute a subspace of
coupling constant space characterized by the constraints~\rf{mcrit}.\
Let $\{g_{i,c}\}$ denote a particular multi-critical point in this
subspace for which the eigenvalues of the matrix model are confined to
the interval $[y_c,x_c]$. If we change the coupling constants we will
move away from the subspace, the $M_k$'s, $k\in[1,m-1]$ will no longer
be zero and the support of the eigenvalue distribution will move to
$[y,x]$. For a variation of the coupling constants of the type
\beq
g_i=g_{i,c}+\delta g_i,\hspace{0.7cm} \delta g_i\sim o(a^m)
\eeq
we will have
\beq
M_k \sim a^{m-k},\hspace{0.7cm}k\in [0,m]
\label{mscal}
\eeq
while the $J$-moments do not scale.
This means that $x$ and $y$ must scale in the following way
\beq
x-x_c\sim a,\hspace{0.7cm}y-y_c\sim a^m.
\eeq
The continuum limit is defined as the limit where we send $a$ to zero
keeping however the string coupling constant $a^{-2m-1}N^{-2}$ fixed.
The definite scaling properties of the moments make these parameters
well suited for studying the continuum limit of the 1-matrix model.
Making use of the scaling properties of the moments
an iterative procedure which allows one
to calculate directly the double scaling relevant versions of $F_g$ and
$W_g(p_1,\ldots,p_s)$ was developed~\cite{hermitian}. The iterative procedure
provided a proof that the continuum relevant version of $F_g$
for a generic model, in the following denoted as $F_g^{(NS)}$,
takes the
following form
\beq
F_g^{(NS)} = \sum_{1< \a_j \leq m} \left\langle \a_1 \ldots \a_s |\a
\right\rangle_g
\frac{M_{\a_1}\ldots M_{\a_s}}{M_1^{\a}d_c^{g-1}}
\hspace{0.7cm} g\geq 1
\hspace{1.0cm} d.s.l.
\label{FNS}
\eeq
where the sum is over all sets of indices obeying the following restrictions
\beq
\sum_{j=1}^{s}\a_j=3g-3+s \hspace{1.0cm} \a=2g-2+s
\eeq
The parameter $d_c$ is the distance between the endpoints of the support
of the eigenvalue distribution, $d_c=x_c-y_c$ and the quantity in
brackets is a real number.
All terms in~\rf{FNS} are
 of the same
order, namely $a^{-(2m+1)(1-g)}$. Bearing in mind the
relation~\rf{mscal} we see that the moments $M_k$ appear as bare
coupling constants or mass parameters of the theory and that continuum
coupling constants, $\mu_k$, can be introduced by
\beq
M_k=a^{m-k}\mu_k,\hspace{0.7cm}k\in[0,m]
\eeq
As in ordinary statistical mechanics the physical coupling constants are
defined not at the critical point but by the {\it approach} to the
critical point. In the neighbourhood of an $m$'th multi-critical point
we can consequently write
\beq
F_g^{(NS)} = \sum_{1<\a_j \leq m} \left\langle \a_1 \ldots \a_s |\a
\right\rangle_g
\frac{\mu_{\a_1}\ldots \mu_{\a_s}}{\mu_1^{\a}}\,\,\,\frac{1}
{{\left[a^{2m+1}d_c\right]}^{g-1}},
\hspace{0.7cm} g\geq 1
\hspace{1.0cm} d.s.l.
\eeq

In the following we will consider a specific type of deformation
$\delta g_i =o(a^m)$ away
from the $m$'th multi-critical point $\{g_{i,c}\}$, namely one for which
$y$ is kept fixed at $y_c$. This choice is not essential but will
simplify our derivations.

\newsection{The Kontsevich model \label{Kontm} }
The Kontsevich model is defined by the partition function
\beq
Z^{Kont}[N,M]=\e^{N^2 F^{Kont}}=
\frac{\int dX \exp\left \{-N\tr \left(\frac{MX^2}{2}
+\frac{iX^3}{6}\right)\right \}}
{\int dX \exp\left \{-N\tr \left(\frac{MX^2}{2}\right)\right\}}
\eeq
where the integration is over $N\times N$ hermitian matrices. This
partition function only depends on the parameters $t_k$
\beq
t_k=\frac{1}{N}\tr M^{-(2k+1)}
\label{tKont}
\eeq
and expressed in terms of these it is a $\tau$-function of the kdV
hierarchy.
As is the case for 1-matrix models the free energy $F^{Kont}$ has a
genus expansion (Cf.\ to equation~\rf{genexp}). The genus-0 contribution
to $F^{Kont}$ reads
\bea
F_0^{Kont}&=&\frac{1}{3}\frac{1}{N}\sum_{i=1}^{N} m_i^3 -
\frac{1}{3}\frac{1}{N}\sum_{i=1}^{N} (m_i^2-2u_0)^{3/2}
-u_0 \frac{1}{N}\sum_{i=1}^N(m_i^2-2u_0)^{1/2} \nonumber \\
&&+ \frac{u_0^3}{6}-\frac{1}{2}\frac{1}{N^2}
\sum_{i,j=1}^N\ln\left\{\frac{(m_i^2-2u_0)^{1/2}+(m_j^2-2u_0)^{1/2}}
{m_i+m_j}\right\}
\label{F0Kont}
\eea
where the $m_i$'s are the eigenvalues of the matrix $M$ and
the parameter $u_0$ is given by the boundary condition
\beq
u_0+\frac{1}{N}\sum_i\frac{1}{\sqrt{m_i^2-2u_0}}=0
\eeq
It can be derived by means of the Dyson Schwinger equations of the model
as done in references~\cite{cm,Kostov}. Al\-ter\-na\-tive\-ly
it can be found by
ex\-ploit\-ing the fact that $Z^{Kont}[N,\{t_k\}]$ is a $\tau$-function of
the kdV hierarchy. The latter approach was followed in
reference~\cite{IZ} where it was also shown that the higher genera
contributions can be written in the following form
\beq
F_g^{Kont} = \sum_{\a_j > 1} \left\langle \a_1 \ldots \a_s |\a
\right\rangle_g^{Kont}\;
\frac{I_{\a_1}\ldots I_{\a_s}}{(1-I_1)^{\a}}
\hspace{0.7cm} g\geq 1
\label{fgkont}
\eeq
where the sum is over sets of indices obeying the following restrictions
\beq
\sum_{j=1}^{s}\a_j=3g-3+s \hspace{1.0cm} a=2g-2+s
\eeq
and where the moments $I_k$ are defined by
\beq
I_k=\frac{1}{N}\sum_{j=1}^{N}\frac{1}{(m_j^2-2u_0)^{k+1/2}}
\hspace{0.7cm}k\geq 0
\eeq
The quantities in brackets can be given an interpretation in terms of
intersection indices on moduli space~\cite{Kon91,Witten}.
The formula~\rf{fgkont} encodes all informations about the kdV hierarchy
and hence all information about the string equations corresponding to
the different multi-critical points mentioned in the previous section.
As shown in reference~\cite{IZ} to obtain the string equation of the
$m$'th multi-critical model from the free energy of the Kontsevich model
one should neglect $I_k$ with $k>m$, keep $I_m$ constant, send
$I_1-1,I_2,\ldots, I_{m-1}$ to zero and introduce a scaling parameter $s$
such that
\beq
v_q=\frac{I_q(1-I_1)^{q-2}}{I_2^{q-1}}
\hspace{0.7cm}3\leq q \leq m
\eeq
remains finite while
\beq
s=\frac{1-I_1}{I_2^{3/5}}
\eeq
tends to zero.
In this limit we hence have
\beq
F_g^{(NS)} = \sum_{1<\a_j \leq m} \left\langle \a_1 \ldots \a_s |\a
\right\rangle_g
\frac{v_{\a_1}\ldots v_{\a_s}}{v_1^{\a}} \frac{1}{s^{5(g-1)}}
\hspace{0.7cm} g\geq 1
\eeq
We see that
this prescription for fine tuning the $I$'s is exactly the
same fine tuning as the one for the $M_k$'s dictated by the double
scaling limit, $s^5$ playing a role similar to that
of $a^{(2m+1)}$. Furthermore it
was shown in reference~\cite{hermitian} that
\beq
\left\langle \a_1 \ldots \a_s |\a \right\rangle_g^{Kont}=
\left\langle \a_1 \ldots \a_s |\a \right\rangle_g
\label{bracket}
\eeq
Hence it is obvious that we have a 1-1 mapping between $F_g^{(NS)}$
and $F_g^{Kont}$ for $g\geq 1$.
However the proof of reference~\cite{hermitian} was based on a somewhat
unconventional limiting procedure as already noted.
  It also left unanswered the question
what parameters of
the original 1-matrix model play the role of continuum time variables.
Here we will show using  the usual double scaling prescription
that it is possible to define continuum time
variables for the generic hermitian matrix model such that the
double scaling limit of its partition function apart from some
complications for genus zero turns into
the partition function of the Kontsevich model.

\newsection{From 1-matrix model to Kontsevich model \label{tau}}
The connecting link between the hermitian 1-matrix model and the
Kontsevich model is the boundary equations of the two models. Below we
first show how these equations immediately tell us how to define
continuum time variables for the hermitian 1-matrix model. The rest of
the section is devoted to proving the following conjecture.
{\it With the definition of continuum times given
in equation~\rf{Tk} the continuum
partition function of the generic hermitian 1-matrix model
deviates from the one of the Kontsevich model only at genus zero.}
The proof is based
on the loop equations of the two models and is yet another example of
the strength of these equations.

\subsection{Definition of continuum time variables.
\label{conttimes} }

To motivate our choice of  continuum time variables $T_k$ for the non
symmetrical hermitian 1-matrix model let us write the boundary equation
for the Kontsevich model as
\beq
\sum_{k=0}^{\infty} c_k\,\, (t_k+\delta_{k,1})\, (2u_0)^k=0, \hspace{1.0cm}
c_k=\frac{(2k-1)!!}{k!\,\, 2^k}
\label{kontbound2}
\eeq
and let us remind the reader that to study the $m$'th kdV flow we should
keep in this equation only $t_0$ and $t_m$.

The idea is now to define $T_k$ in such a way that by taking the double
scaling limit of the boundary equations~\rf{B1} and~\rf{B2} we
reproduce equation~\rf{kontbound2} with the $T_k$'s replacing the
$t_k$'s.
Here and in the following we will consider a 1-matrix model for which
the eigenvalues, at the critical point, are confined to the interval
$[y_c,x_c]$.
As mentioned earlier, when we move away from a given $m$'th
multi-critical point by a change of coupling constants $\delta g_i\sim
o(a^m)$ in general both $x$ and $y$ will change. For the simplicity of
the presentation we restrict ourselves to the subclass of
deformations for which $y$ is kept
fixed at $y_c$.
Expanding equation~\rf{B1} in powers of $(x-x_c)$ we find
\beq
\sum_{p=0}^{\infty}c_p(x-x_c)^p M_p^c\left(\{g_i\}\right)=0
\label{B1exp}
\eeq
where
\beq
M_p^c\left(\{g_i\}\right)=M_p\left(x_c,y_c,\{g_i\}\right),
\eeq
Note that the coupling constants entering $M_p^c$ are completely
arbitrary. Rewriting the boundary equation~\rf{B2} we find
\beq
\sum_{p=0}^{\infty}c_p(x-x_c)^p
\left\{x_c\,M_p^c\left(\{g_i\}\right)+M_{p-1}^c\left(\{g_i\}\right)\right\}
= 2
\label{B2exp}
\eeq
Since $M_p^c\sim a^{m-p}$, for $0\leq k\leq m$ the term $M_{p-1}^c$ is
subleading when compared to $M_p$ except for $M_{-1}$ which is equal to
2 up to subleading terms of $o(a^{m+1})$. If we set
\beq
x-x_c=a\,(2u_0) \label{scalxu}
\eeq
and define our continuum time variables by
\beq
T_k+\delta_{k,1} =a^{k+1/2}\, d_c^{1/2}\,
 M_k^c\left(\{g_i\}\right) \hspace{0.7cm} k\geq 0
\label{Tk}\\
\eeq
both~\rf{B1exp} and~\rf{B2exp} turn into the boundary equation of the
Kontsevich model~\rf{kontbound2}\footnote{
We note that the subclass of deformations considered above does not include
the deformations that are associated with $m$'th  multi-critical
behaviour in the usual 1-matrix model sense. These deformations correspond
to the situation where only $T_m$ and $T_0$ are different from zero.
Such a situation can only be obtained if we scale the coupling constants
as  $g_i=g\cdot g_{i,c}$ which
requires that both $x$ and $y$ must scale. However the arguments of
this and all following sections can be repeated for the case where $y$
is not kept fixed. The expressions just become more involved.}
(The reason why we include an additional factor
$\sqrt{a}\,\,d_c^{1/2}$ will become clear later.)

\subsection{Continuum loop equations for the 1-matrix model}
In view of the scaling behaviour~\rf{scalxu} and the general form of
$W_g(p)$ it is natural to introduce continuum momentum variables $\pi_i$
by
\beq
p_i=x_c+a\pi_i
\label{scalp}
\eeq
We define continuum correlators $W^{cont}(\pi_1,\ldots,\pi_s)$ by
\bea
W(p)-\frac{1}{2}V'(p)& \stackrel{d.s.l.}\longrightarrow& \frac{1}{a}
\left(W^{cont}(\pi)-\frac{1}{2}\sum_{k=0}^{\infty}
 (T_k+\delta_{k,1}) \pi^{k-1/2} \right)
\label{W1cont}\\
W(p_1,p_2) +\frac{1}{2}\frac{1}{(p_1-p_2)^2}
&\stackrel{d.s.l.}\longrightarrow
&\frac{1}{a^2}\left(W^{cont}(\pi_1,\pi_2)+
\frac{\frac{1}{2}(\pi_1+\pi_2)}{2(\pi_1-\pi_2)^2\sqrt{\pi_1\pi_2}}\right)
\label{W2cont} \\
W(p_1,\ldots,p_s)&\stackrel{d.s.l.}\longrightarrow & \frac{1}{a^s}\,
W^{cont}(\pi_1,\ldots,\pi_s)
\label{Wscont}
\eea
The subtraction needed to make contact with continuum physics only
concerns the genus zero contribution to the 1- and 2-loop correlators.
This is a well known feature of the double scaling limit.
In particular we have
\beq
W(p,p)\stackrel{d.s.l.}\longrightarrow \frac{1}{a^2}
\left(W^{cont}(\pi,\pi)+\frac{1}{16\pi^2}\right)
\eeq
We note that the prescription~\rf{scalp} for taking the double scaling
limit implies that in this limit the cut $[y,x]$ on the real axis for
$W(p)$ is replaced by a cut of the type $[-\infty,2u_0]$. This is of
course a pleasing result since this is exactly the analyticity structure
of the 1-loop correlator of the Kontsevich model~\cite{cm}.
Let us comment on the $a$ factors extracted on the right hand side. As
shown in reference~\cite{hermitian} in the double scaling limit the loop
insertion operator reduces to
\beq
\frac{d}{dV(p)} = \sum_{n=1}^{\infty}
\frac{dM_n}{dV(p)}\frac{\partial}{\partial M_n}
+\frac{dx}{dV(p)} \frac{\partial}{\partial x} \hspace{0.7cm} d.s.l
\label{lio1dsl}
\eeq
where
\bea
\frac{dM_n}{dV(p)}&=&
                   -(n+1/2)\left\{d_c^{-1/2}(p-x)^{-n-3/2}
                   -M_{n+1}\frac{dx}{dV(p)}\right\}\,
\hspace{0.7cm} d.s.l.
\label{loi2dsl}\\
\frac{dx}{dV(p)}&=&\frac{1}{M_1}d_c^{-1/2}(p-x)^{-3/2}
\label{lio3dsl} \hspace{0.7cm} d.s.l.
\eea
Now expanding the moment $M_p$ in powers of $x-x_c$ gives
\beq
M_p=\sum_{l=p}^{\infty} a^{-p-1/2}\, (T_l+\delta_{1,l})\,\, (2u_0)^{l-p}
\frac{\Gamma(l+1/2)}{\Gamma(p+1/2)\,\,(l-p)!},
\label{momexp}
\hspace{0.7cm}p\geq 1
\eeq
(From this expression one can read off the continuum scaling behaviour of
the moments for a given $m$'th multi-critical model and the
relation~\rf{mscal} is easily reproduced.)
Using equation~\rf{momexp} the scaling relation~\rf{scalxu} for $x$ and the
boundary equation expressed in terms of $T_k$'s one can by means of the
chain rule show that the following relation holds
\beq
\frac{d}{dV(p)}\stackrel{d.s.l.} \longrightarrow \,\,
\frac{1}{a}\,
\frac{d}{dV^{cont}(\pi)} = \frac{1}{a}
\left\{-\sum_{k=0}^{\infty}(k+1/2)\frac{1}{\pi^{k+3/2}}\frac{d}{dT_k}\right\}
\label{dVdsl}
\eeq
Bearing in mind the relation~\rf{loopaverage} it appears natural to extract one
power of $a^{-1}$ for each loop in a given correlator. The
relation~\rf{dVdsl} will be essential for the proof of our conjecture.
Due to the peculiarities of the genus zero contributions to the 1- and
2-loop correlators it is convenient to write the loop equations in a
genus expanded version. Inserting the genus expansion~\rf{Wgen}
into~\rf{loopeqn} it is seen that $W_g(p), \,\,g\geq 1$ obeys the
following equation.
\beq
\left\{\hat{K}-2W_0(p)\right\}W_g(p)=
\sum_{g'=1}^{g-1}W_{g'}(p)\,W_{g-g'}(p)+W_{g-1}(p,p),\hspace{0.7cm} g\geq 1
\eeq
where
\beq
\hat{K}f(p)=\cim\frac{V'(\om)}{p-\om} f(\om)
\eeq
Let us now introduce in this equation the continuum correlators. First
we note that since $W_0(p)$ itself does not scale we have in the double
scaling limit
\beq
\left\{\hat{K}-2W_0(p)\right\}W_g(p) \stackrel{d.s.l.} \longrightarrow
\,\, \frac{1}{a^2}\,\cim \frac{V'(\om)-2W_0(\om)}{p-\om}W_g^{cont}(\om)
\eeq
Using the definition~\rf{W1cont} the right hand side (times $a^2$)
can be written as
\bea
rhs &=& -\left\{2W_0^{cont}(\pi)
      -\sum_k (T_k+\delta_{k,1})\,\pi^{k-1/2}\right\} W_g^{cont}(\pi)
\nonumber\\
&&      -\oint_{\infty}\frac{d\om}{2\pi i}
      \left\{\frac{2W_0^{cont}(\om)-\sum_k(
         T_k+\delta_{k,1})\om^{k-1/2}}{\pi-\om}
\right\}
W_g^{cont}(\om)
\label{rhs}
\eea
where $\oint_\infty$ denotes a contour integral where the contour
encircles infinity. To proceed let us write $W_g^{cont}(\om)$ as
\beq
W_g^{cont}(\om)=\sum_{q=0}^{\infty} \om^{-q-3/2}W_g^{cont,q}
\label{Wgexp}
\eeq
That $W_g^{cont}$ allows an expansion of this type is obvious for $g\geq
1$ since for $g\geq 1$ we have
$W_g^{cont}(\om)=\frac{d}{dV^{cont}(\om)}F_g$, where
$\frac{d}{dV^{cont}(\om)}$ is given by equation~\rf{dVdsl}. (It is
also evident from the explicit expression for $W_g(p)$ given in
reference~\cite{hermitian}.) That the same is true for $g=0$ will become
clear in section~\rf{proof}. Performing the contour integral in~\rf{rhs}
and making use of the definition~\rf{W2cont} one obtains the following
continuum loop equation
\bea
&&-\left[2W_0^{cont}(\pi)-\sum_k (T_k+\delta_{k,1})\pi^{k-1/2}\right]
W_g^{cont}(\pi)- \sum_{q,a}(T_{2+a+q}+\delta_{2+q+a,1})W_g^{cont,q}\, \pi^a
 \nonumber \\
&&\hspace*{+0.6cm}=\sum_{g'=1}^{g-1}W_g^{cont}(\pi)W_{g-g'}^{cont}(\pi)
+ W_{g-1}^{cont}(\pi,\pi)
+\delta_{g,1}\cdot \frac{1}{16\pi^2}
\label{loopcont}
\eea

\subsection{Loop equations for the Kontsevich model}
Inspired by the equation~\rf{dVdsl} let us introduce a loop insertion
operator for the Kontsevich model by
\beq
\frac{d}{dV^{Kont}(\pi)} \equiv
-\sum_k(k+1/2)\frac{1}{\pi^{k+3/2}}\frac{d}{dt_k}
\label{dVKont}
\eeq
and multi-loop correlators by
\beq
W^{Kont}(\pi_1,\ldots,\pi_s)=\frac{d}{dV^{Kont}(\pi_s)}\ldots
\frac{d}{dV^{Kont}(\pi_1)} F^{Kont},\hspace{0.7cm} s\geq 1
\eeq
Using the relation~\rf{tKont} it is easy to show by means of the
chain rule that
\beq
\frac{d}{dV^{Kont}(\pi)}= N\left.\frac{\partial}{\partial
m_i^2}\right|_{m_i^2=\pi}
\eeq
In this form the loop insertion operator can readily be applied
to~\rf{F0Kont} to yield
\beq
 W_0^{Kont}(\pi)=\frac{1}{2}\left[
-\sqrt{\pi-2u_0}+\sqrt{\pi}+\frac{1}{N}\sum_{j=1}^{N}\frac{1}{\pi-m_j^2}
\left(\frac{\sqrt{\pi-2u_0}}{\sqrt{m_j^2-2u_0}}-\frac{m_j}{\sqrt{\pi}}\right)
\right]
\label{W1Kont}
\eeq
and
\beq
W_0^{Kont}(\pi_1,\pi_2)=\frac{1}{4(\pi_1-\pi_2)^2}
\left\{\frac{\pi_1+\pi_2-4u_0}{\sqrt{\pi_1-2u_0}\sqrt{\pi_2-2u_0}}
-\frac{\pi_1+\pi_2}{\sqrt{\pi_1\,\pi_2}}\right\}
\label{W2Kont}
\eeq
With the definitions given above the master equation of the Kontsevich
model can be written as
\beq
\frac{1}{N^2}\left\{W^{Kont}(\pi,\pi)+\frac{1}{16\pi^2}\right\}
+\left(\tilde{W}^{Kont}(\pi)\right)^2
+\frac{1}{N}\sum_{j=1}^{N}\frac{\tilde{W}^{Kont}(m_j^2)}{m_j^2-\pi}
=\frac{\pi}{4}
\eeq
where
\beq
\tilde{W}^{Kont}(\pi)=W^{Kont}(\pi)
-\frac{1}{2}\sum_k(t_k+\delta_{k,1})\pi^{k-1/2}
\eeq
To make contact with the previous section we will rewrite this equation
in a genus expanded version. To deal with the sum appearing on the left
hand side above let us note that the genus $g$ contribution to the
1-loop correlator can be expanded in the following way
\beq
W_g^{Kont}(\pi)=\sum_{q=0}^{\infty}\pi^{-q-3/2}W_g^{Kont,q}
\eeq
Then making use of the definition of the $t_k$'s,~\rf{tKont} we arrive
at the following form of the loop equation.
\bea
&&-\left[2W_0^{Kont}(\pi)-\sum_k (t_k+\delta_{k,1})\pi^{k-1/2}\right]
W_g^{Kont}(\pi)- \sum_{q,a}(t_{2+a+q}+\delta_{2+q+a,1})W_g^{Kont,q}\, \pi^a
 \nonumber \\
&&\hspace*{+0.6cm}=\sum_{g'=1}^{g-1}W_g^{Kont}(\pi)W_{g-g'}^{Kont}(\pi)
+ W_{g-1}^{Kont}(\pi,\pi)
+\delta_{g,1}\cdot \frac{1}{16\pi^2}
\label{loopKont}
\eea
\subsection{Proof of our conjecture \label{proof}}
The task of proving our conjecture amounts to proving the following two
identities
\bea
W_0^{cont}(\pi)&=&W_0^{Kont}(\pi) \label{W1Kc}\\
W_0^{cont}(\pi_1,\pi_2)&=&W_0^{Kont}(\pi_1,\pi_2) \label{W2Kc}
\eea
where it is understood that the quantities on the left hand side should
be expressed in terms of $T_k$'s whereas those on the right hand side
should be expressed in terms of $t_k$'s. By comparing
equation~\rf{loopcont} and~\rf{loopKont} it is easily seen that once
this task has been fulfilled it follows by induction that
\beq
W_g^{cont}(\pi,\{T_k\})=W_g^{Kont}(\pi,\{t_k\}), \hspace{0.7cm}g\geq 1
\label{W1gKc}
\eeq
since now the two sets of loop equations and corresponding boundary
equations only differ by $\{T_k\}$ appearing in one case and $\{t_k\}$
in the other. Furthermore by taking a glance at equation~\rf{dVdsl}
and~\rf{dVKont} bearing in mind the relations~\rf{loopaverage}
and~\rf{W1cont} -- \rf{Wscont} one easily convinces oneself that
\beq
W_g^{cont}(\pi_1,\ldots,\pi_s,\{T_k\}) =
W_g^{Kont}(\pi_1,\ldots,\pi_s,\{t_k\}), \hspace{0.7cm}g,s \geq 1
\eeq
{}From equation~\rf{W1cont} and~\rf{W1gKc} it  follows that
\beq
W_g(p)\stackrel{d.s.l.}\longrightarrow
\frac{1}{a}W_g^{Kont}(\pi),\hspace{0.7cm} g \geq 1
\eeq
since both the second term on the left hand side and the second term on
the right hand side of~\rf{W1cont} are of zeroth order in genus. Now due to the
similarity between $\frac{d}{dV^{cont}(\pi)}$ and
$\frac{d}{dV^{Kont}(\pi)}$ we immediately find
\beq
F_g \stackrel{d.s.l.} \longrightarrow F_g^{Kont}, \hspace{0.7cm} g\geq 1
\label{Fgcont}
\eeq
To address the $g=0$ case we note that equation~\rf{W2cont}
and~\rf{W2Kc} imply that the double scaling limit of $W_0(p_1,p_2)$
differs from $W_0^{Kont}(\pi_1,\pi_2)$ only by a term which does not
depend on any couplings. Therefore we have
\beq
W_0^{cont}(\pi_1,\ldots,\pi_s)=
W_0^{Kont}(\pi_1,\ldots,\pi_s), \hspace{0.7cm} s\geq 3
\eeq
However, equation~\rf{W1Kc} does not allow us to conclude anything about
the relation between the genus zero contributions to the partition
functions of the two models because of the subtractions appearing in
equation~\rf{W1cont}.

Let us now turn to the proof of the relations~\rf{W1Kc} and~\rf{W2Kc}.
The proof of the latter is by far the most straightforward since
$W_0(p_1,p_2)$ is universal, i.e. it does not contain any explicit
reference to the coupling constants.
By acting with $\frac{d}{dV(p_2)}$ on $W_0(p_1)$ one gets
\beq
W_0(p_1,p_2)=
\frac{1}{2(p_1-p_2)^2}
\left\{\frac{p_1p_2-\frac{1}{2}(p_1+p_2)(x+y)+xy}
{\sqrt{(p_1-x)(p_1-y)(p_2-x)(p_2-y)}}-1 \right\}
\eeq
and taking the double scaling limit following the
prescriptions~\rf{scalxu} and~\rf{scalp}
one easily reproduces~\rf{W2cont} with $W_0^{Kont}(\pi_1,\pi_2)$
replacing $W_0^{cont}(\pi_1,\pi_2)$.
To prove the relation~\rf{W1Kc} we will prove that
\beq
W_0(p)-\frac{1}{2}V'(p)+
\frac{1}{a}\frac{1}{2}\sum_{k=0}^{\infty}(T_k+\delta_{k,1})\pi^{k-1/2}
\,\,\stackrel{d.s.l.}\longrightarrow \,\,\frac{1}{a}\,W_0^{Kont}(\pi)
\label{altproof}
\eeq
First we note that due to equation~\rf{W0} we can write the two first
terms as
\beq
W_0(p)-\frac{1}{2}V'(p)=\frac{1}{2}\,\oint_{\infty}\frac{d\om}{2\pi i}
\frac{V'(\om)}{p-\om}\left\{\frac{(p-x)(p-y)}{(\om-x)(\om-y)}\right\}^{1/2}
\label{W0sub}
\eeq
Furthermore inserting our definition of continuum times~\rf{Tk}
into the remaining term on the left hand side of~\rf{altproof}
we get
\bea
\lefteqn{\frac{1}{a}\frac{1}{2}\sum_{k=0}^{\infty}(T_k+\delta_{k,1})\pi^{k-1/2}
 =\frac{1}{2}\sum_{k=0}^{\infty}(a\pi)^{k-1/2}\,
d_c^{1/2}\,M_k^c\left(\{g_i\}\right)
}\nonumber \\
&=&
-\left\{\left.W_0(p)-\frac{1}{2}V'(p)\right\}\right|_{x-x_c,y=y_c}
+\frac{1}{2}(p-x_c)^{-1/2}(p-y_c)^{1/2}
M_0^c\left(\{g_i\}\right)
\nonumber
\\
&=&-\frac{1}{2}\oint_{\infty}\frac{d\om}{2\pi i}\frac{V'(\om)}{p-\om}
\left\{\frac{(p-y_c)(\om-x_c)}{(p-x_c)(\om-y_c)}\right\}^{1/2}
\label{Trew}
\nonumber \\
\eea
To obtain the second equality sign we have made use of the rewriting of
$W_0(p)$ given in~\rf{W0exp} and the scaling relation for
$p$,~\rf{scalp}. (In particular we have used that $(p-y)$ in the
continuum limit can be replaced by $d_c$.) We note that it was in order
to be able to carry out this step that we had to multiply our boundary
equation \rf{B1exp} by $d_c^{1/2}\sqrt{a}$.
To obtain the third equality sign we
have made use of the relation~\rf{W0sub}. So our statement is now the
following
\beq
W_0^{Kont}=\lim_{d.s.l}\,a\cdot
\left[\frac{1}{2}\oint_{\infty}\frac{d\om}{2\pi i}
\frac{V'(\om)}{p-\om}\left\{
\frac{\sqrt{(p-x)(p-y_c)}}{\sqrt{(\om-x)(\om-y_c)}}-
\frac{\sqrt{(\om-x_c)(p-y_c)}}{\sqrt{(p-x_c)(\om-y_c)}}\right\}
\right]
\label{Wfin}
\eeq
The similarity with equation~\rf{W1Kont} is striking and the equality is
straightforward to prove. To do so one expands $(p-\om)^{-1}$ in powers
of $\left(\frac{p-x_c}{\om-x_c}\right)$ and the quantities $(p-x)$ and
$(\om-x)$ in powers of $\left(\frac{x-x_c}{p-x_c}\right)$ and
$\left(\frac{x-x_c}{\om-x_c}\right)$ respectively. The factor $(p-y_c)$
can simply be replaced by $d_c$. The $\frac{1}{\sqrt{\pi}}$ term of~\rf{Wfin}
vanishes as it should. This is actually ensured by the boundary
equation. In the process of
expanding the integrand it proves convenient to pull out a factor
$(p-x)^{1/2}$. The result of the expansion procedure is the following
expression for the right hand side of~\rf{Wfin}
\bea
rhs&=& \lim_{d.s.l.} \left\{
\frac{1}{2}(p-x)^{1/2}\,(p-y_c)^{1/2}
 \sum_{b=1}^{\infty}\sum_{m=1}^{b-1}c_b\,\,
\frac{(x-x_c)^b}{(p-x_c)^{b-m+1}}M_m^c\right\}
 \\
&=& \frac{1}{2}(\pi-2u_0)^{1/2}\sum_{b=1}^{\infty}\sum_{m=1}^{b-1}
c_b \frac{(2u_0)^b}{\pi^{b-m+1}}(T_m+\delta_{m,1})
\label{rhsW}
\eea
By rewriting~\rf{W1Kont} using the definition~\rf{tKont} of times it is easy to
show that the expression~\rf{rhsW} is exactly $W_0^{Kont}(\pi)$.

\newsection{Discussion \label{discussion}}
We have in the present paper considered the generic
hermitian matrix model. However the same strategy can be applied to the
complex matrix model. The complex matrix model is in all respects very
similar to the hermitian matrix model. It has a set of loop equations
which can be written in the same form as that of equation~\rf{loopeqn}. The
appropriate requirement concerning the analyticity structure of its
1-loop correlator is that it has only one square root branch cut
$[-\sqrt{z},\sqrt{z}]$
 on the real axis. With this requirement one can solve the
loop equations genus by genus. The solution of course depends on the
parameter $z$  which is determined by a boundary condition similar
to~\rf{B2}. As before the multi-loop correlators can be found by
applying a loop insertion operator to the 1-loop correlator and as
before expressing the higher genera contributions to the correlators is
facilitated by introducing a moment description.
To relate the double scaling of the partition function of the complex
matrix model to the one of the Kontsevich model one takes the same line
of action as for the hermitian matrix model. Appropriate continuum time
variables are defined by the requirement that the boundary equation of
the complex matrix model reproduces the boundary equation of the
Kontsevich model when the double scaling limit is taken.
The resulting variables turn out to be related to the moments of the complex
matrix model by an equation similar to~\rf{Tk}. Furthermore
for the loop insertion operator one has again a relation
like~\rf{dVdsl}. However, a closer analysis of the loop equations shows
that in stead of~\rf{Fgcont} we have
\beq
F_g^C \stackrel{d.s.l.}\longrightarrow \frac{1}{4^{g-1}}
\left(2F_g^{Kont}\right),\hspace{0.7cm}g\geq 1
\eeq
and that in the double scaling limit the partition function of
the complex matrix model involving matrices of size $N\times N$ turn
into the square of the partition function of the Kontsevich model
involving matrices of size $2N \times 2N$ except for some complications
at genus zero.  The same is true for the
continuum partition function of the reduced hermitian matrix model since
in reference~\cite{hermitian} it was shown that
\beq
F_g^C=\frac{1}{4^{g-1}}F_g^{(S)}\hspace{0.7cm} d.s.l.
\eeq
where $F_g^{(S)}$ is the genus $g$ contribution to the free energy of
the reduced model. (We refer to~\cite{hermitian} for
details.) Let us mention that the factor two in difference between
the double scaling limit of the free energy of the symmetrical and the
generic hermitian matrix model is easy to understand in the eigenvalue
picture. It arises because the double scaling
relevant part of $F_g$ in the case of a symmetrical model gets
contributions from both ends of the support of the eigenvalue
distribution while in the generic case there is only critical
behaviour associated with one end of the eigenvalue distribution. And
let us stress that it is not possible to pass from the generic
to the symmetrical case after the continuum limit has been
taken since the possibility of having different behaviour at the two
endpoints exists only in the generic case.
\vspace{12pt}
\newline
\noindent
{\bf Acknowledgements}\hspace{0.3cm}
We thank L. Chekhov,
Yu.\ Makeenko, A.\ Marshakov and A.\ Mironov for interesting
and valuable discussions.

\end{document}